\documentclass[traditabstract]{aa} 
\usepackage{txfonts} 
\usepackage[english]{babel} 
\usepackage{t1enc} 
\usepackage{graphicx} 
\usepackage{subfig} 
\usepackage{lscape} 
\usepackage{natbib} 
\usepackage{footmisc} 
\usepackage{url} 
\usepackage{dingbat}

\begin{document} 
 
\title{Trigonometric Parallaxes of Massive Star Forming Regions:
G012.88+0.48 and W33}                                                    
\author{K. Immer\thanks{Member of the International Max Planck Research
School (IMPRS) for Astronomy and Astrophysics at the Universities
of Bonn and Cologne}\inst{1,2} \and M.~J. Reid\inst{2} \and K.~M. Menten\inst{1} 
\and A. Brunthaler\inst{1} \and T.~M. Dame\inst{2}}                                             
 
\institute{ 
Max-Planck-Institut f\"ur Radioastronomie, Auf dem H\"ugel 69, D-53121
Bonn, Germany                                                           
\and 
Harvard-Smithsonian Center for Astrophysics, 60 Garden Street, 02140,
Cambridge, MA, USA}                                                     
\date{Received xxxx; accepted xxxx} 
 
\abstract{We report trigonometric parallaxes for water masers in the
G012.88+0.48 region and in the massive star forming complex W33 
(containing G012.68--0.18, G012.81--0.19, G012.90--0.24,
G012.90--0.26), from the Bar and Spiral Structure Legacy (BeSSeL) survey using the Very Long
Baseline Array. The parallax distances to all these masers are consistent with
$2.40^{+0.17}_{-0.15}$ kpc, which locates the W33 complex and G012.88+0.48
in the Scutum spiral arm. Our results show that
W33 is a single star forming complex at about two-thirds the kinematic
distance of 3.7 kpc.  The luminosity and mass of this region, 
based on the kinematic distance, have therefore been overestimated by 
more than a factor of two. The spectral types in the star cluster in W33\,Main 
have to be changed by 1.5 points to later types.}

 
\authorrunning{K. Immer et al.} 
\titlerunning{Trigonometric Parallaxes of G012.88+0.48 and W33} 
 
\maketitle 
 
\section{Introduction} 
 
\begin{figure*}[htbp] 
     \centering 
      \includegraphics[width=18cm]{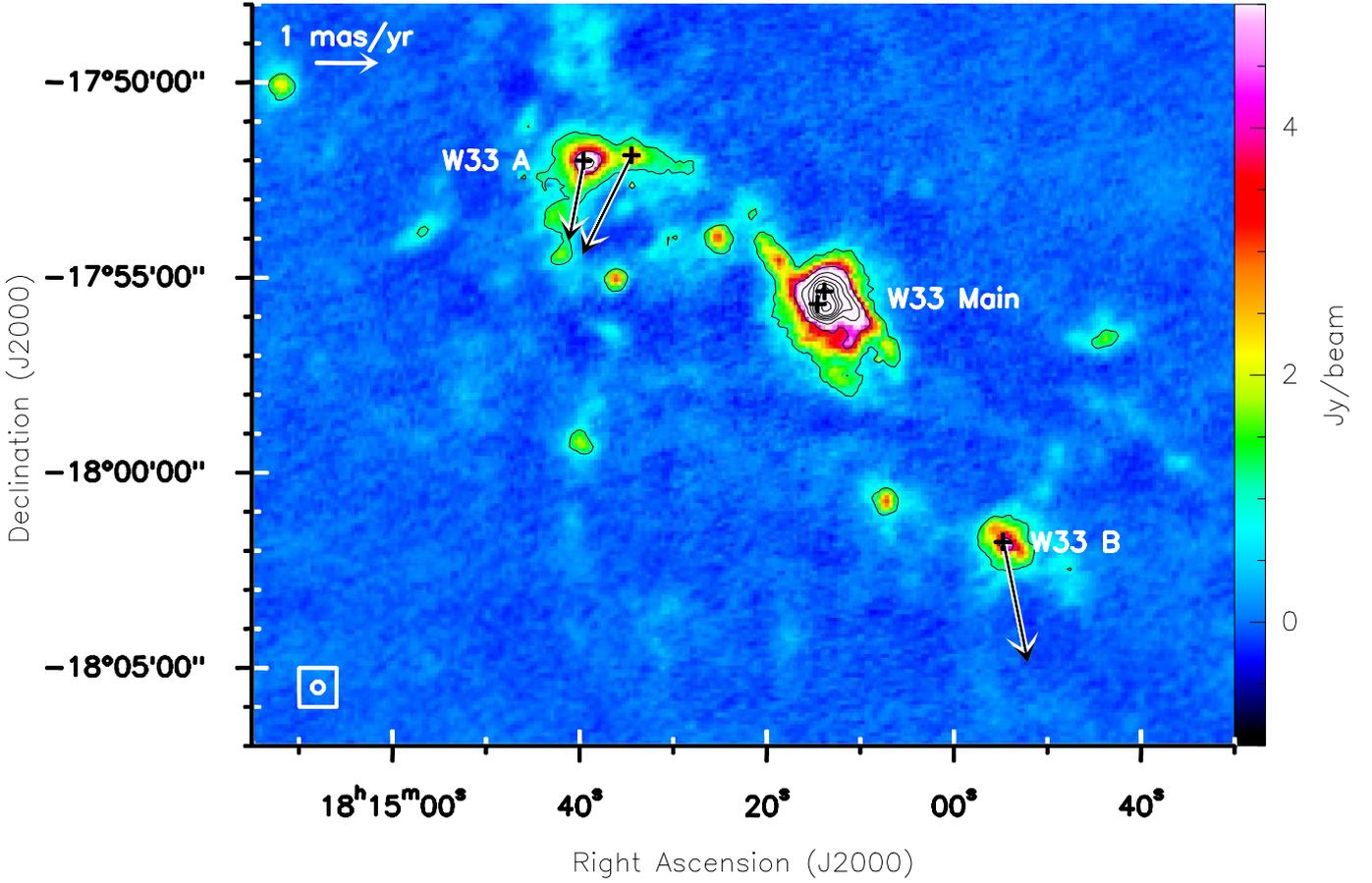} 
   \caption{870 $\mu$m dust emission of the W33 complex from the
ATLASGAL survey with contour levels of 1, 4, 7, 10, 13, 16, 20, and 30 Jy beam$^{-1}$. 
The positions of the water masers are indicated with black crosses. The arrows
show the internal motion vectors of the two clouds W33\,A and W33\,B 
in the reference frame of W33\,Main. The length of the arrow in the 
upper left corner corresponds to a motion vector of 1 mas yr$^{-1}$. 
The circle in the lower left corner shows the beam of the ATLASGAL 
observations. G012.88+0.48 is located at an angular distance of about 
0.7 degrees to the north of W33, outside of this figure.}                                 
   \label{W33_ATLASGAL} 
\end{figure*}

W33 is a massive and very luminous star forming complex that contains
typical star forming regions from quiescent infrared dark clouds
to highly active, infrared bright clouds, associated with \ion{H}{ii} regions. 
Observations of the dust emission in the W33 complex from the APEX 
Telescope Large Area Survey of the GALaxy \citep[ATLASGAL,][]{Schuller2009} 
at 870 $\mu$m show three large molecular clouds and several smaller dust clumps
(Fig. \ref{W33_ATLASGAL}). Each of the large clouds contains water
masers at G012.68--0.18 (W33\,B), G012.81--0.19 (W33\,Main), G012.90--0.24
(W33\,A), and G012.90--0.26 (W33\,A) \citep{Genzel1977, Jaffe1981}.                    
  
To derive fundamental physical parameters of the molecular clouds
such as their size, luminosity, mass, and the spectral types of embedded
stars, we have to know their distances. One commonly used method is
to dermine the kinematic distance of a cloud from radial velocity
measurements. Comparing the radial velocity of the cloud with a rotation
curve of the Galaxy \citep[e.g.][]{Burton1978, Reid2009b} yields
the Galactocentric radius from which the distance of the cloud can
be determined. However, in the first and fourth quadrants there are 
two possible distances corresponding to a given radius, complicating
the distance determination. In addition, local velocity deviations
due to shocks, outflows or noncircular motions can lead to large
errors of the kinematic distance \citep{Xu2006}.                                       
 
Kinematic distance determination for the W33 complex is complicated by a
peculiar kinematic structure that was discovered via H$_2$CO absorption
line and radio recombination line observations \citep{Bieging1978,
Goss1978, Bieging1982}. Two radial velocity components at 36 and
58 km s$^{-1}$ were detected over extended parts of the region. The
first component is concentrated in the northeastern part
(W33\,A) and the second component peaks in
the southwestern part (W33\,B). This velocity structure
can be explained by either one connected star forming region with
large internal motions at a near kinematic distance of 3.7 kpc (corresponding
to the radial velocity of 36 km s$^{-1}$), or a superposition of
independent star forming regions along the line
of sight \citep{Haschick1983, Goss1978, Gardner1975, Bieging1978}.      
 
Assuming a near kinematic distance of 3.7 kpc to W33, the angular
size of the complex of 15$\arcmin$ can be converted into a physical
size of 16 pc. \citet{Stier1984} determined a total infrared luminosity
of $\sim$2$\cdot$10$^6$ L$_{\sun}$ for W33 at this distance. 
Using $^{13}$CO maps and a typical value of 
n($^{13}$CO)/n(H$_{2}$) = 4 $\cdot$ 10$^{-7}$, \citet{Goldsmith1983}
estimated the total mass of the complex to be between 0.2 and 2 
$\cdot$10$^6$ M$_{\sun}$.
From radio continuum observations, \citet{Haschick1983} inferred the
existence of an OB star cluster in W33\,Main with individual spectral types 
ranging from O6 to B0, based on a kinematic distance of 4 kpc. 

W33\,A is a well-studied star forming region. The source 
has a high far-infrared luminosity of 10$^{5}$ L$_{\sun}$ \citep{Faundez2004}
at the previously assumed kinematic distance of 3.7 kpc.
\citet{GalvanMadrid2010} detected parsec-scale filaments of 
cold molecular gas around two dust cores. The interaction of 
these filaments might have triggered the star formation 
activity in the cores. The brightest core drives a strong outflow and 
the dynamics in this core show evidence for a rotating disk, 
perpendicular to the outflow. These results are supported by CO observations of 
\citet{Davies2010} which also suggest the presence of a rotationally-flattened 
cool molecular envelope. Br$\gamma$ emission in W33\,A seems to 
trace a fast bipolar wind which has the same orientation 
as the large-scale outflow \citep{Davies2010}.  
The colder and more massive second core in W33\,A seems to be in 
an earlier evolutionary stage, driving a more modest outflow 
\citep{GalvanMadrid2010}.

G012.88+0.48 is a high-mass protostellar object \citep{Sridharan2002,
Beuther2002a} which is prominent for its strong maser emission from the 
OH, H$_2$O, and CH$_3$OH molecules.                                
G012.88+0.48 has a radial velocity of 33.8 km s$^{-1}$ \citep{Sridharan2002},
locating the star forming region at a distance of 3.6 kpc, assuming it is at 
its near-kinematic distance. This soure is projected about 0.7 degrees from 
the main star forming regions of W33.
\citet{Sridharan2002} and \citet{Beuther2002a} determined
the luminosity and mass of this star forming region to be L $\sim$ 3$\cdot$10$^5$
L$_{\sun}$ and M $\sim$ 3$\cdot$10$^3$ M$_{\sun}$ at this distance.
\citet{Xu2011} determined the trigonometric
parallax for the 12.2 GHz methanol masers in this source. They obtained
a value of 0.428$\pm$0.022 mas which corresponds to a distance
of 2.34 kpc, revising the kinematic distance by a factor of 0.65 (= 2.34 kpc/3.6 kpc). 
If G012.88+0.48 is associated with the W33 complex, then the kinematic 
distances of the complex are too large. Alternatively, it is possible that 
G012.88+0.48 is not associated with W33 and is substantially closer to the Sun.

Other recent trigonometric parallax measurements have shown that
kinematic distance measurements can be wrong by more than a factor
of two \citep[e.g.][]{Reid2009b, Sato2010a}. Trigonometric parallax
observations of maser sources, however, have proven to be a reliable
method to accurately measure distances to star forming regions whose
accuracy is limited only by observational uncertainties.                
 
The aim of this project is to obtain distances to the water masers
in the W33 complex from trigonometric parallax observations and,
therefore, address the question of whether the molecular clouds in W33 belong
spatially to one connected star forming complex or are at different 
distances and projected near each other on the sky.  Also, solid
distances will provide the
basis for more accurate values for the luminosity and mass of the
complex. 
 
The observations presented in this paper are part of the Bar and Spiral 
Structure Legacy (BeSSeL) 
Survey\footnote{\url{http://bessel.vlbi-astrometry.org/index.shtml}}
\citep{Brunthaler2011}, a  National Radio Astronomy 
Observatory's\footnote{The National 
Radio Astronomy Observatory (NRAO) is a facility of the 
National Science Foundation operated under cooperative agreement by 
Associated Universities, Inc.} Very Long Baseline Array (VLBA) key
project which will determine the distances to hundreds of star forming
regions. We report results of the parallax and proper motion measurements
toward the water masers in the sources G012.88+0.48 and the massive
star forming complex W33.

\section{Observation and Data Reduction} 
\label{Obs}  
 
The massive star forming region G012.88+0.48 and the W33
complex (G012.68--0.18, G012.81--0.19, G012.90--0.24, G012.90--0.26)
were observed in the $6_{16}-5_{23}$ transition of the H$_2$O molecule
(rest frequency 22.23508 GHz), using the VLBA. The general observation
setup and data calibration procedures are described in \citet{Reid2009a}.
Here, we only describe the details specific to the present sources.
 
All water masers were observed at nine epochs (VLBA program BR145I)
with a total observing time per epoch of $\sim$~7 hours. 
The observing dates (2010 Sep 19, 2010 Oct 22, 2010 Dec 19, 
2011 Feb 15, 2011 Mar 26, 2011 May 21, 2011 Jul 30, 2011 Sep 20, 
2012 Jan 03) were selected to well sample the
Right Ascension parallax signature in time, since the Declination
parallax signature is much smaller. However, due to bad weather
conditions, the data of the first epoch could not be used for the
parallax measurements.                                               
 
Three different quasars, J1808--1822, J1809--1520, and J1825--1718,
from the VLBA calibrator search for the BeSSeL survey \citep{Immer2011}
and the ICRF2 catalog \citep{Fey2009}, were observed for background
position references. The background source J1809--1520 could only
be detected in the first four epochs, and thus was not used for the
parallax determination. Four adjacent frequency
bands with 4 MHz bandwidth each were used in right and left circular
polarization. The maser signals were centered in the second band
at an LSR velocity, V$_{LSR}$, of 45 km s$^{-1}$ for all masers.
The channel spacing of the observations is 0.42 km s$^{-1}$.            
 
In order to correct for instrumental
phase offsets between the frequency bands, observations of the calibrator
J1800+3848 from the ICRF2 catalog were used. The strongest maser
feature in the water maser G012.68--0.18 at a V$_{LSR}$ of 59.3 km
s$^{-1}$ was used for the phase-referencing of the data.                
 
The water masers G012.81--0.19, G012.90--0.24, G012.90--0.26, and G012.88+0.48
were much weaker than G012.68--0.18 and were calibrated using G012.68--0.18
as phase reference. This yields positions relative to G012.68--0.18, 
from which relative parallaxes can be estimated.                                   

For the data reduction, we used the software Parseltongue
\footnote{http://www.radionet-eu.org/rnwiki/ParselTongue} \citep{Kettenis2006}, a 
scripting interface for NRAO's Astronomical Image Processing System (AIPS).
After calibrating the data, the maser emission and the continuum
sources were imaged with the AIPS task IMAGR with a circular Clean beam
of 2 mas for all epochs. The positions of the masers and the background
sources J1808--1822 and J1825--1718 were then determined by fitting
Gaussian brightness distributions to the images \citep[see][]{Reid2009a}.
Absolute positions of the strongest maser spot in G12.68--0.81 were
derived relative to the two background sources. For the remaining
water masers, relative positions to G012.68--0.18 were determined.
To obtain absolute parallaxes, we added the positions of G012.68--0.18
relative to both background quasars to the position measurements
of these water masers.                                                  
 
The positions were modeled with a sinusoidal parallax signature and
linear proper motions in each coordinate. Since systematic errors 
(typically from unmodeled atmospheric delays) are normally much 
larger than random noise, the formal position errors from
the Gaussian fits are mostly too small to account for the total error
of the observations which leads to high $\chi^2$ values in the modeling
process.  The magnitude of the systematic errors can
only be inferred from the quality of the parallax fit. Thus, we defined
"error-floors" for the Right Ascension and Declination maser positions
and added those in quadrature to the Gaussian fit errors. The "error-floor"
values were adjusted until the $\chi^2$ value per degree of freedom was close to unity
for both coordinates \citep[see][]{Reid2009a}.
 
In order to determine the internal gas kinematics in each source, we measured
the proper motions for all maser spots that were detected in at least
three epochs without solving for the parallax parameter.

\section{Results} 
\label{Results}

\begin{table*} 
\begin{minipage}[t]{\textwidth} 
 \caption{Coordinates, line-of-sight velocities, and peak flux densities 
of the background quasars and the strongest maser spots.}   
 \label{ObsTab} 
  \centering 
\renewcommand{\footnoterule}{} 
  \begin{tabular}{ccccccc}\hline 
Source & R.A. (J2000) & Dec. (J2000) & V$_{LSR}$ & F$_{Peak}$ & rms
& Cloud Association\\                                                   
            & [hh mm ss.ssss] & [ dd '' ""."""] & [km s$^{-1}$] & [Jy
beam$^{-1}$] & [Jy beam$^{-1}$] &\\ \hline                             
J1808--1822       &  18 08 55.5150 & $-$18 22 53.383   & & 0.012 & 0.001 &\\ 
J1809--1520       &  18 09 10.2094 & $-$15 20 09.699   & & 0.022 & 0.001 &\\ 
J1825--1718       &  18 25 36.5323 & $-$17 18 49.848  &  & 0.096 & 0.001 &\\ 
G012.68--0.18    &  18 13 54.7457 & $-$18 01 46.588   & 59.3 &132.5 &
0.001 & W33\,B\\                                                         
G012.81--0.19    &  18 14 13.8283 &  $-$17 55 21.035  & 34.1/$-$1.4 & 1.0/0.2
& 0.018/0.007 & W33\,Main\\                                              
G012.90--0.24    &  18 14 34.4366 & $-$17 51 51.891   & 34.9 & 18.3
& 0.081 & W33\,A\\                                                       
G012.90--0.26    &  18 14 39.5714  & $-$17 52 00.382    & 37.0 & 3.0
& 0.016 & W33\,A\\                                                       
G012.88+0.48     &  18 11 51.4939  &  $-$17 31 28.911  & 29.4 & 23.3&
0.133 &\\                                                               
   \end{tabular}
\tablefoot{
Columns 2, 3, and 4 list the coordinates and line-of-sight velocities of the 
background quasars and the maser spots that were used for parallax fitting. 
The peak flux densities of the background quasars and the strongest maser 
spots and the rms of the reference channel images, obtained from epoch 
2011 May 21, are listed in Cols. 5 and 6. The last column shows with which 
cloud the masers in the W33 complex are associated.
}
\end{minipage} 
\end{table*} 

\begin{table*}
\begin{minipage}[t]{\textwidth} 
 \caption{Parallax and proper motion results.}                                            
 \label{ParallaxResults} 
  \centering 
  \begin{tabular}{lccccccc}\hline 
\multicolumn{1}{c}{Maser}  & V$_{LSR, maser}$ & V$_{LSR, thermal}$\tablefootmark{a} & Parallax $\pi$ & Distance d & $\mu_{x}$ & $\mu_{y}$ \\                                                  
\multicolumn{1}{c}{Source} & km s$^{-1}$ & km s$^{-1}$ & [mas] & [kpc] & [mas yr$^{-1}$] & [mas
yr$^{-1}$]\\ \hline          
\vspace{1mm} G012.68--0.18  (W33\,B)   & 59.3 & $\sim$55 & 0.416$\pm$0.028 & 2.40$^{+0.17}_{-0.15}$ & $-$1.00$\pm$0.30 & $-$2.85$\pm$0.29\\               
\vspace{1mm} G012.81--0.19 (W33\,Main) & $-$1.4 & $\sim$36 & 0.343$\pm$0.037 & 2.92$^{+0.35}_{-0.28}$ & $-$0.24$\pm$0.17 & +0.54$\pm$0.12 \\           
\vspace{1mm} G012.81--0.19 (W33\,Main)    & 34.1 & $\sim$36 & 0.343$\pm$0.037 & 2.92$^{+0.35}_{-0.28}$ & $-$0.60$\pm$0.11 & $-$0.99$\pm$0.13\\            
\vspace{1mm} G012.90--0.24 (W33\,A)   & 34.9 & $\sim$36 &0.408$\pm$0.025 & 2.45$^{+0.16}_{-0.14}$ &  +0.19$\pm$0.08 & $-$2.52$\pm$0.32\\             
\vspace{1mm} G012.90--0.26 (W33\,A)   & 37.0 & $\sim$37 & 0.396$\pm$0.032 & 2.53$^{+0.22}_{-0.19}$& $-$0.36$\pm$0.08 & $-$2.22$\pm$0.13\\ \hline       
\vspace{1mm} G012.88+0.48     & 29.4 & $\sim$34& 0.340$\pm$0.036 & 2.94$^{+0.35}_{-0.28}$ 
& +0.12$\pm$0.13 & $-$2.66$\pm$0.23\\             
   \end{tabular} 
\tablefoot{
\tablefoottext{a}{\citet{Sridharan2002, Purcell2012, Wienen2012}.}
Columns 6 and 7 give the proper motions in eastward and northward directions. 
The proper motions of the first maser are obtained from
fitting a model of expanding outflows to all maser spots. The proper motions
of the remaining sources are averages of the proper motions of all
maser spots in each source.
}
\end{minipage} 
\end{table*} 

Table \ref{ObsTab} gives the coordinates (Cols. 2, 3), line-of-sight velocities (Col. 4), 
and peak flux densities from epoch 2011 May 21 (Col. 5) of the background quasars 
and the maser spots that were used for the parallax fits. Column 6 lists the rms brightness of 
the reference channel images, obtained from epoch 2011 May 21. In the last column, 
we show the associations of the masers in the W33 complex.

The parallax and proper motion results of all masers are summarized
in Table \ref{ParallaxResults}. Columns 2, 3, and 4 give the line-of-sight velocities of the
masing and thermal gas and the parallaxes of the maser sources. In the sixth and seventh
columns, the proper motions in eastward and northward directions
are listed.

One goal of our observations is to estimate the distance and 
proper motion of the central object 
in each source. In G012.68--0.18, we obtained the proper motion of the central object 
by fitting a model for expanding outflows
\citep{Sato2010b} to the proper motions of all maser spots. In the other sources, 
the small number of maser spots and their distribution did not 
allow the fitting of this model to their proper motions. Instead, the presented values are 
averages of the proper motions of all maser spots in each source.

\subsection{W33}
 
\begin{figure}[htbp] 
     \centering 
      \includegraphics[width=5cm]{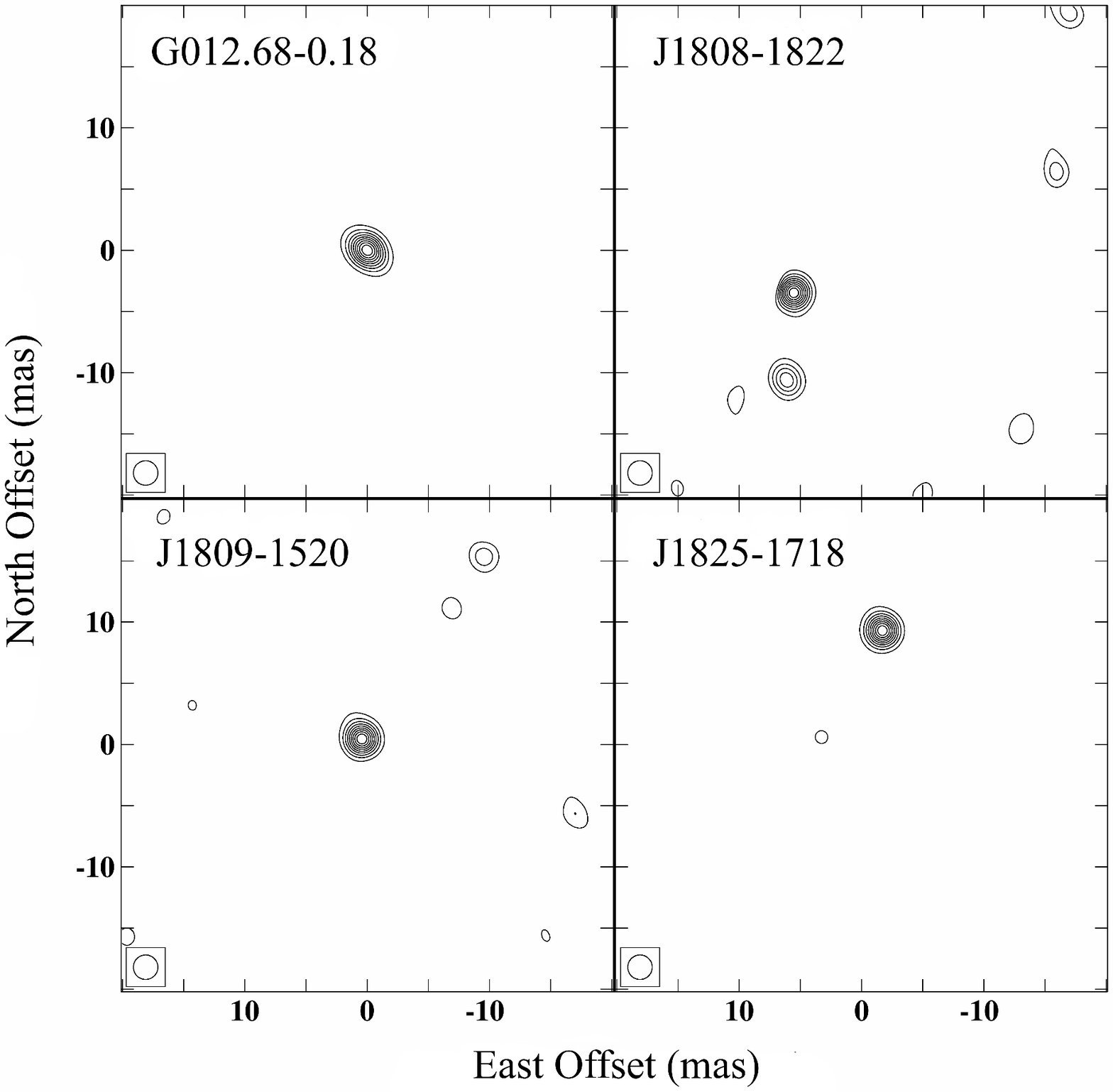}
   \caption{Left upper panel: Maser emission in reference channel
(59.3 km s$^{-1}$). Right upper, left and right lower panel: Compact
emission of the three background quasars J1808--1822, J1809--1520,
and J1825--1718, which were phase-referenced to the reference maser
channel. The images are obtained from epoch 2011 May 21. The contour
levels are 10 to 100 \% in steps of 10 \% of the maximum in each
panel (G012.68--0.18: 132.5 Jy beam$^{-1}$, J1808--1822: 12.3 mJy beam$^{-1}$,
J1809--1520: 22.1 mJy beam$^{-1}$, J1825--1718: 96.1 mJy beam$^{-1}$).
The restoring beam is shown in the lower left corner.}                  
   \label{G12.68} 
\end{figure}

\begin{figure}[htbp] 
     \centering 
     \includegraphics[width=9cm]{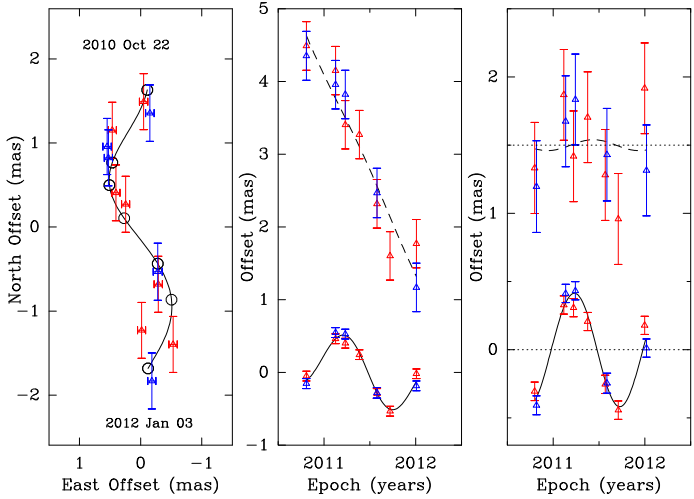} 
        \caption{Parallax and proper motion fits of G012.68--0.18,
yielding a parallax of (0.416$\pm$0.028) mas. The positions of
the reference maser spot at 59.3 (triangles) km s$^{-1}$ were measured
relative to the background quasars J1808--1822 (blue) and J1825--1718
(red). Left Panel: Position on the sky with a label for the first
and last epoch. Middle Panel: East (continuous line) and North (dashed
line) position offsets with parallax and proper motion fits versus
time. Right Panel: Right ascension (continuous line) and Declination
(dashed line) parallax fits with the best-fit proper motions removed,
showing only the parallax signature.}                                   
   \label{G12.68_ParallaxFit} 
\end{figure} 
 
\begin{figure} 
	\centering 
      \includegraphics[width=9cm]{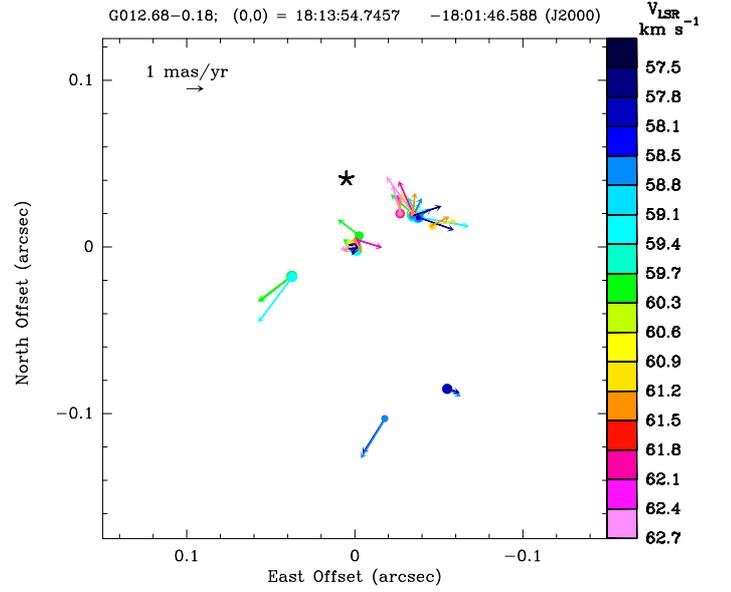}
       \caption{Positions of all maser spots in G012.68--0.18 that
have been detected in at least three epochs (positions from epoch
2011 May 21). The size of the spots scales logarithmically with the flux
density of the spots. The asterisk marks the position of the central star which
has an error of $\sim$0.1$\arcsec$ in both coordinates. The arrows
show the proper motions of the maser spots in the reference frame
of the central star.}                                  
	\label{G12spotmap} 
\end{figure}

\begin{figure}[htbp] 
     \centering 
     \includegraphics[width=9cm]{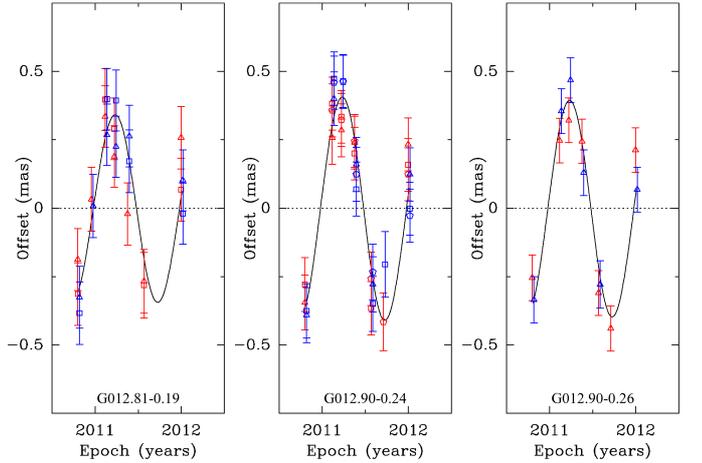} 
        \caption{Right ascension parallax fits of G012.81--0.19 (left),
G012.90--0.24 (middle), and G012.90--0.26 (right), yielding absolute
parallaxes of $0.343 \pm 0.037$ mas, $0.408 \pm 0.025$ mas, and
$0.396 \pm 0.032$ mas, respectively. The positions of the maser spots 
were measured relative to the water maser G012.68--0.18
and consecutively to the two background quasars J1808--1822 (blue)
and J1825--1718 (red). Left panel: Positions of the maser
spots at 34.1 (triangles) and $-$1.4 (squares) km s$^{-1}$. Middle
panel: Positions of the maser spots at 34.5 (triangles),
34.9 (squares), and 35.3 (pentagons) km s$^{-1}$. Right panel: Positions
of the maser spot at 37.0 (triangles) km s$^{-1}$.}          
   \label{G12.81_ParallaxFit} 
\end{figure} 
 
\begin{figure} 
	\centering 
       \includegraphics[width=9cm]{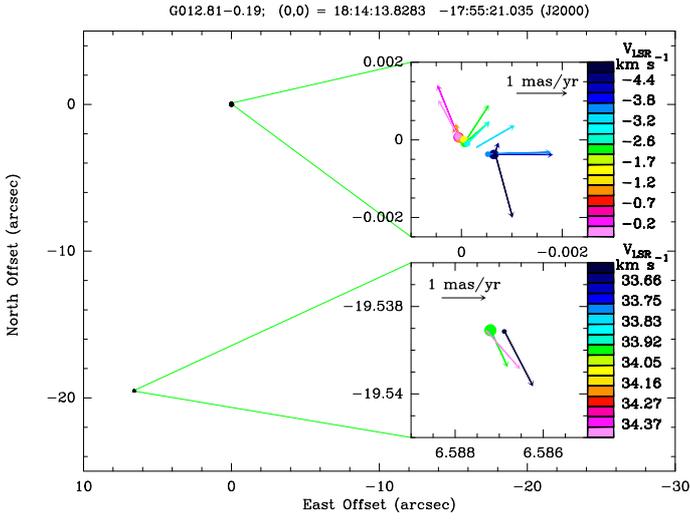}
       \caption{Left panel: Positions of all maser spots in G012.81--0.19
that have been detected in at least three epochs (positions from
epoch 2011 May 21). Inside panels: Zoom-in on the two maser spot groups
at 34 km s$^{-1}$ (lower panel) and $-$1 km s$^{-1}$ (upper panel). The maser spots 
in the upper panel are ordered along a ridge. 
The size of the spots scales logarithmically with the flux density of the
spots. The arrows show the absolute proper motions of the maser spots.}      
	\label{G12.81spotmap} 
\end{figure}

\begin{figure} 
	\centering 
       \includegraphics[width=9cm]{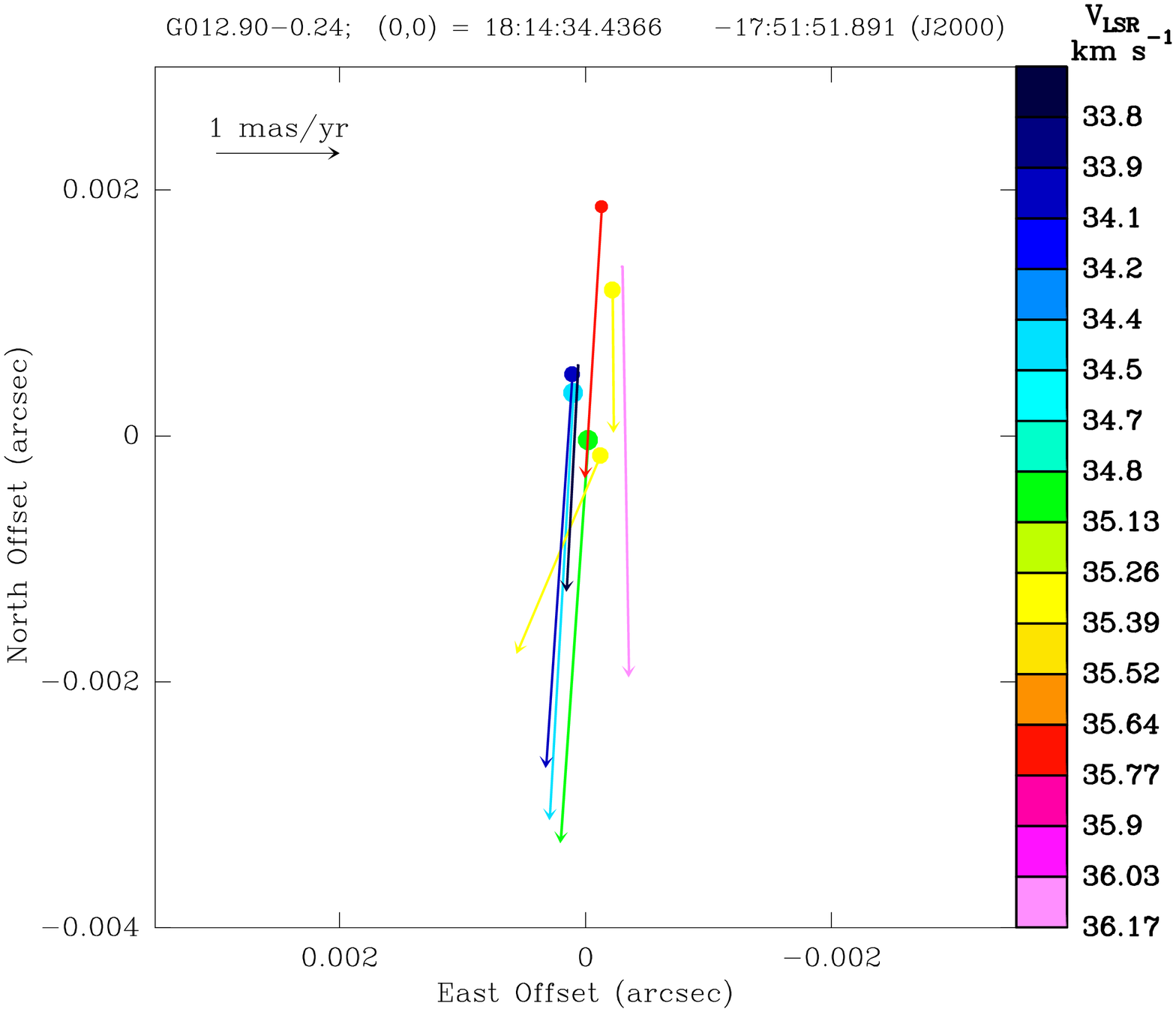} 
       \caption{Positions of all maser spots in G012.90--0.24 that
have been detected in at least three epochs (positions from epoch
2011 May 21). The size of the spots scales logarithmically with the flux
density of the spots. The arrows show the absolute proper motions of the maser spots.}                                                            
       \label{G12.90spotmap} 
\end{figure} 
 
\begin{figure} 
	\centering 
       \includegraphics[width=9cm]{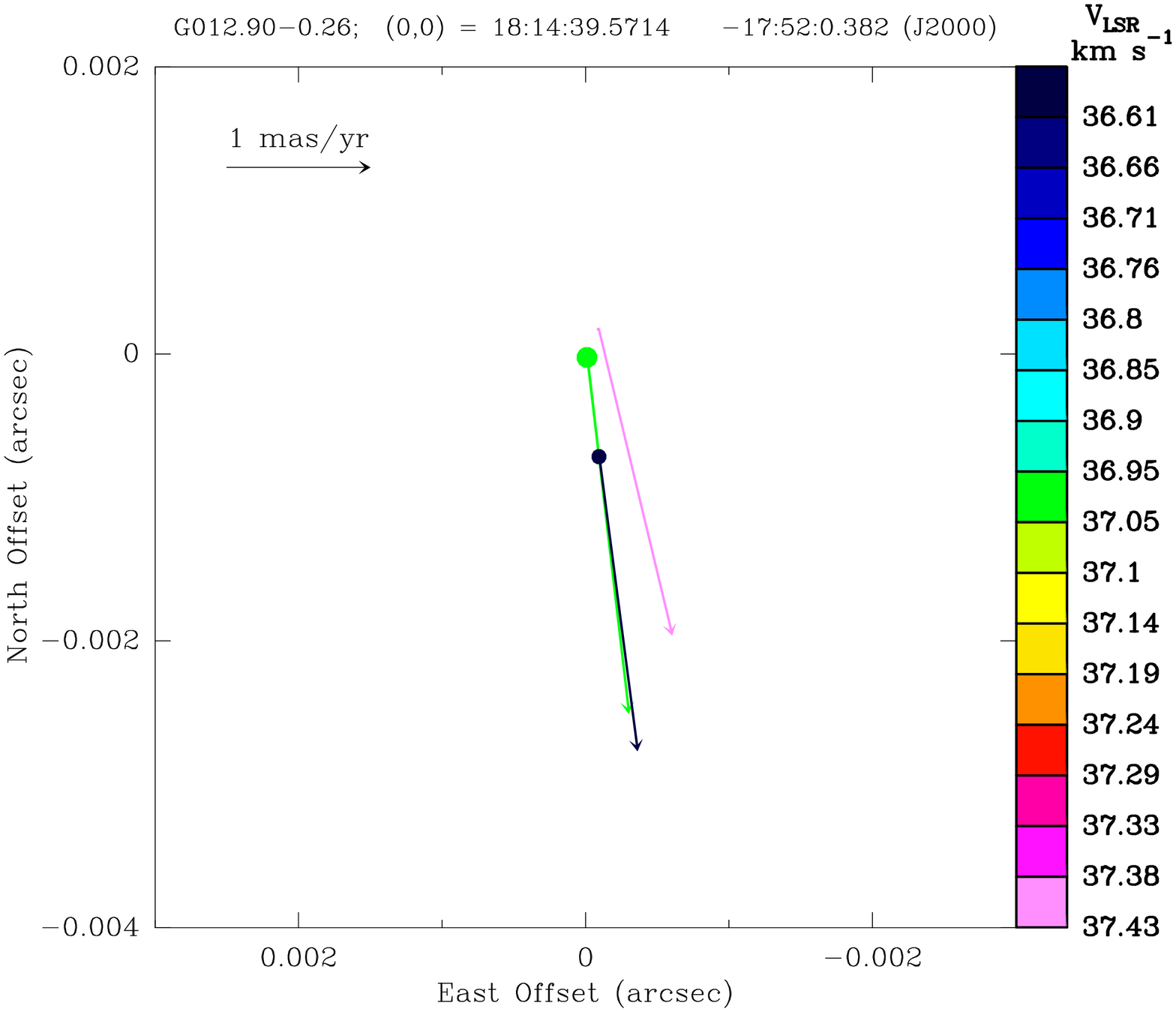} 
       \caption{Positions of all maser spots in G012.90--0.26 that
have been detected in at least three epochs (positions from epoch
2011 May 21). The size of the spots scales logarithmically with the flux
density of the spots. The arrows show the absolute proper motions of the maser spots.}                                                            
	\label{G12.91spotmap} 
\end{figure} 
  
Figure \ref{W33_ATLASGAL} shows the dust emission of the W33 complex
from the ATLASGAL survey at 870 $\mu$m. The three most massive 
clouds W33\,A, W33\,Main, and W33\,B are indicated. The water masers 
are marked with black crosses.    
 
Water maser spots were detected in an $0{\rlap.}$\,$''$\,$\! 15 \times
0{\rlap.}$\,$''$\,$\! 15$ area in W33\,B (G012.68--0.18), covering a velocity
range of 57 -- 63 km s$^{-1}$. Figure \ref{G12.68} shows that the maser
emission in the reference channel ($V_{LSR} = 59.3$ km s$^{-1}$,
left panel) as well as the emission of the three background quasars
J1808--1822 (right upper panel), J1809--1520 (left lower panel),
and J1825--1718 (right lower panel) are dominated by one compact
component. Since the quasar J1809--1520 was detected only in the
first four epochs, we determined the position of the reference maser
spot only relative to the background quasars J1808--1822 and J1825--1718
as a function of time (see Fig. \ref{G12.68_ParallaxFit}). For three
epochs (2010 Oct 22, 2010 Dec 19, 2011 Sep 20), we used a different
maser spot at the same velocity as position reference since the original
reference maser spot, used in the other epochs, was too weak to serve as a
phase reference. 

 
The positions of the reference maser spot in G012.68--0.18 
relative to background quasars were fitted
with the sinusoidal parallax signature and a linear proper
motion. We obtained parallax estimates of $\pi=0.462 \pm 0.034$
mas and $\pi=0.387 \pm 0.037$ mas, relative to the
background quasars J1808--1822 and J1825--1718, respectively. Since
both results are consistent within their joint uncertainty, 
we combined the two data sets
and repeated the fitting, yielding a value of $\pi=0.416 \pm 0.028$ mas 
which corresponds to a distance of $2.40^{+0.17}_{-0.15}$ kpc 
(see Fig. \ref{G12.68_ParallaxFit}).                                
 
We fitted the proper motions of all maser spots in G012.68--0.18, 
detected in at least three epochs, with a model for expanding outflows with
the position and proper motion of the central driving source as
free parameters \citep[for more details about the model see][]{Sato2010b}. 
The proper motion of the central star is 
$\mu_{x}=-1.00 \pm 0.30$ mas yr$^{-1}$ and $\mu_{y}=
-2.85\pm0.29$ mas yr$^{-1}$. 
Figure \ref{G12spotmap} shows the positions of
the maser spots with their proper motions in the reference frame
of the central source. Observations with the Submillimeter Array (SMA) 
at 230 GHz (Immer et al, in prep.) show that the reference maser spot is 
located in a dust core, offset by $\sim$ 0.1 $\arcsec$ from the submillimeter 
continuum peak. The radial velocity of the reference maser spot is 
similar to the systemic velocity of the thermal gas in this cloud 
\citep[55 km s$^{-1}$,][]{Wienen2012}.                                              
 
The remaining W33 masers and G012.88+0.48 were phase-referenced to
G12.68-0.18 and the positions of their strongest maser spots were determined
relative to this maser, yielding relative parallaxes. To obtain absolute
parallax values, the positions of G012.68--0.18 relative to the two
background quasars were added to the position measurements of the
other water masers.                                                     
 
In G012.81--0.19, we detected two spatially and kinematically distinct
maser groups, covering velocity ranges of 33 to 35 km s$^{-1}$ and
$-$5 to 0 km s$^{-1}$, separated by $\sim$21$\arcsec$. 
The positions of two maser spots at $-$1.4 and 34.1 km s$^{-1}$ were determined
relative to G012.68--0.18. Combining the two position measurements,
we obtained an absolute parallax of $\pi = 0.343 \pm 0.037$ mas
(see Fig. \ref{G12.81_ParallaxFit}, left panel). The positions of
the two maser groups are shown in Fig. \ref{G12.81spotmap}. The images
on the right-hand side show both maser groups on a smaller scale
with their proper motions. Averaging the proper motions of all maser
spots in each group, yields a proper motion value for each velocity
group (uncorrected for expansion): $\mu_{x, 34 \textnormal{km s$^{-1}$}} = -0.24 
\pm 0.17$ mas yr$^{-1}$ and $\mu_{y, 34 \textnormal{km s$^{-1}$}} = 
0.54 \pm 0.12$ mas yr$^{-1}$ as well as $\mu_{x, -1 \textnormal{km s$^{-1}$}} = 
-0.60 \pm 0.11$ mas yr$^{-1}$ and $\mu_{y, -1 \textnormal{km s$^{-1}$}} = 
-0.99 \pm 0.13$ mas yr$^{-1}$. The maser spots in the velocity 
group around $-$1 km s$^{-1}$ are ordered along a ridge with redshifted 
velocities at the north-east end and blueshifted velocities at the south-west 
end. The maser spot group at 34 km s$^{-1}$ is located in a dust core, 
offset by $\sim$ 7 $\arcsec$ from the main continuum emission peak at 230 GHz 
(Immer et al., in prep.). The radial velocity of the reference maser spot is similar 
to the systemic velocity of the thermal gas of 36 km s$^{-1}$, which was 
estimated from ammonia observations of the H$_{2}$O southern Galactic Plane Survey 
\citep[HOPS][]{Purcell2012}. 
The maser spot group at $-$1 km s$^{-1}$ is associated with emission from a dust 
core at 345 GHz (Immer et al., in prep.). However, the radial velocities of these 
maser spots are very different from the systemic velocity in this source and they 
might trace the blueshifted shock front of an outflow.

The maser emission from G012.90--0.24 spans a velocity range of 33 to
37 km s$^{-1}$. For the parallax fitting, we used the positions of
three maser spots at velocities of 34.5, 34.9, and 35.3 km s$^{-1}$
and obtained an absolute parallax of $\pi = 0.408 \pm 0.025$
mas (see Fig. \ref{G12.81_ParallaxFit}, middle panel). The proper
motions of the detected maser spots are very similar (Fig. \ref{G12.90spotmap}).
The proper motion of the source (uncorrected for expansion), obtained
from averaging all maser spot proper motions, is $\mu_{x} = 
0.19 \pm 0.08$ mas yr$^{-1}$ and $\mu_{y} = -2.52 \pm 0.32$ mas
yr$^{-1}$. The ATLASGAL map (Fig. \ref{W33_ATLASGAL}) shows that 
the reference maser spots are offset by $\sim$ 8 $\arcsec$ from a submillimeter 
continuum peak. The radial velocities of the reference maser spots are close to 
the systemic velocity of G012.90--0.24 (36 km s$^{-1}$, estimated from 
the HOPS survey).                                                              
 
Only three maser spots were detected in G012.90--0.26, covering a
velocity range of 36 to 38 km s$^{-1}$. We used the strongest maser
spot at 37.0 km s$^{-1}$ for the parallax fitting, yielding an absolute 
parallax of $\pi = 0.396 \pm 0.032$ mas (see
Fig. \ref{G12.81_ParallaxFit}, right panel). Averaging the proper
motions of the three maser spots, we obtain a proper motion of
the source of $\mu_{x} = -0.36 \pm 0.08$ mas yr$^{-1}$ and
$\mu_{y} = -2.22 \pm 0.13$ mas yr$^{-1}$.
The position of the reference maser spot is consistent with 
the submillimeter continuum peak MM1-SE, detected by \citet{GalvanMadrid2010} 
at 230 GHz. This object is associated with a faint radio source at 7 mm 
\citep{vanderTak2005}. The radial velocity of the reference maser spot is very close 
to the systemic velocity of the thermal gas in W33\,A 
\citep[38.5 km s$^{-1}$,][]{GalvanMadrid2010}. A comparison of the 
proper motion of G012.90--0.26 with 
CO observations of the large-scale outflow in W33\,A \citep{GalvanMadrid2010} 
show that their directions do not agree. Thus, we will assume that the proper motion 
of this water maser is not strongly influenced by the outflow but probably 
reflects the motion of the central object.                  
 
\subsection{G012.88+0.48}

\begin{figure}[htbp] 
     \centering 
     \includegraphics[width=9cm]{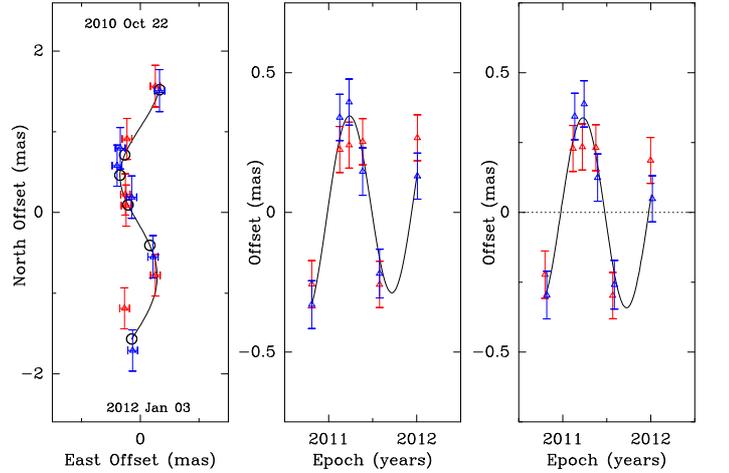} 
        \caption{Parallax and proper motion fits of G012.88+0.48,
yielding an absolute parallax of $0.340 \pm 0.036$ mas. The positions
of the maser spot at 29.4 (triangles) km s$^{-1}$ were
measured relative to the water maser G012.68--0.18 and consecutively
to the two background quasars J1808--1822 (blue) and J1825--1718
(red). Left Panel: Position on the sky with a label for the first
and last epoch. Middle Panel: East position offsets with parallax
and proper motion fits versus time. Right Panel: Right ascension
parallax fit with the best-fit proper motions removed, showing only
the parallax signature.}                                                
   \label{G12.89_ParallaxFit} 
\end{figure} 
 
\begin{figure} 
	\centering 
\includegraphics[width=9cm]{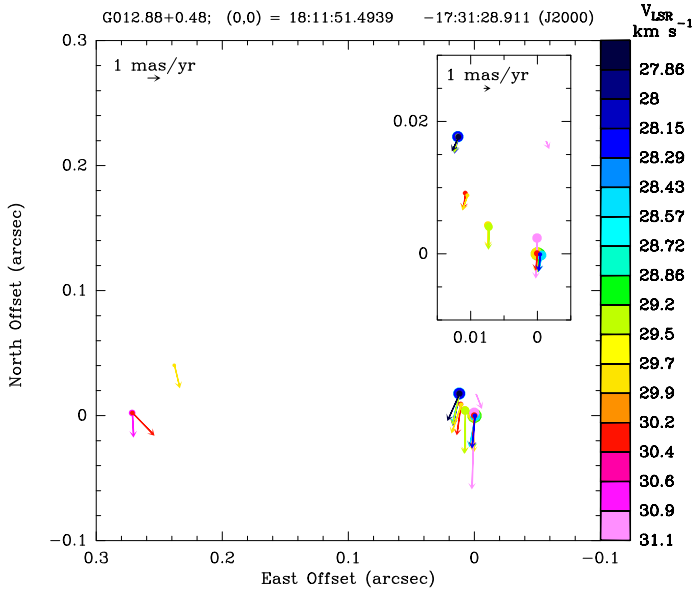}
       \caption{Positions of all maser spots in G012.88+0.48 that
have been detected in at least three epochs (positions from epoch
2011 May 21). The size of the spots scales logarithmically with the flux
density of the spots. The arrows show the absolute proper 
motions of the maser spots.
The smaller picture is a zoom on the western group of maser spots. 
These spots might be tracing a bow shock in this source.}                                                            
	\label{G12.89spotmap} 
\end{figure}

Maser emission in the spectrum of G012.88+0.48 was detected only in
the velocity range 27 to 31 km s$^{-1}$. Fitting the positions of
the strongest maser spot at 29.4 km s$^{-1}$, we obtained an absolute
parallax of $\pi = 0.340 \pm 0.036$ mas (see Fig. \ref{G12.89_ParallaxFit}),
which is consistent within 2$\sigma$ (joint uncertainty)
with the parallax result of \citet{Xu2011} for
12.2 GHz methanol masers in this star forming region.                   
 
Figure \ref{G12.89spotmap} shows two groups of maser spots, separated by
$\sim$ 0.25$\arcsec$. The spatial distribution of the spots in 
the western group shows an arc-like morphology. The combination with 
the proper motions of these spots seems to indicate that the water masers 
trace a shock front in this source. Methanol maser observations in G012.88+0.48 
at 6.7 GHz with the Japanese VLBI Network and the East-Asian VLBI Network 
present a similar picture with a group of maser spots arranged in an arc-like 
structure with line-of-sight velocities
from 30 to 40 km s$^{-1}$ \citep[][Fujisawa et al., in prep.]{Fujisawa2012}. 
This maser group is offset from our western group by $-$0.3$\arcsec$ 
and $-$1.2$\arcsec$ in east-west and north-south direction, respectively.

We averaged the proper motions of all water maser
spots to obtain one proper motion for G012.88+0.48, yielding $\mu_{x}
= 0.12 \pm 0.13$ mas yr$^{-1}$ and $\mu_{y} = -2.66 
\pm 0.23$ mas yr$^{-1}$.                                                                                           
 
\section{Discussion}

\begin{figure} 
	\centering 
       \includegraphics[width=10cm]{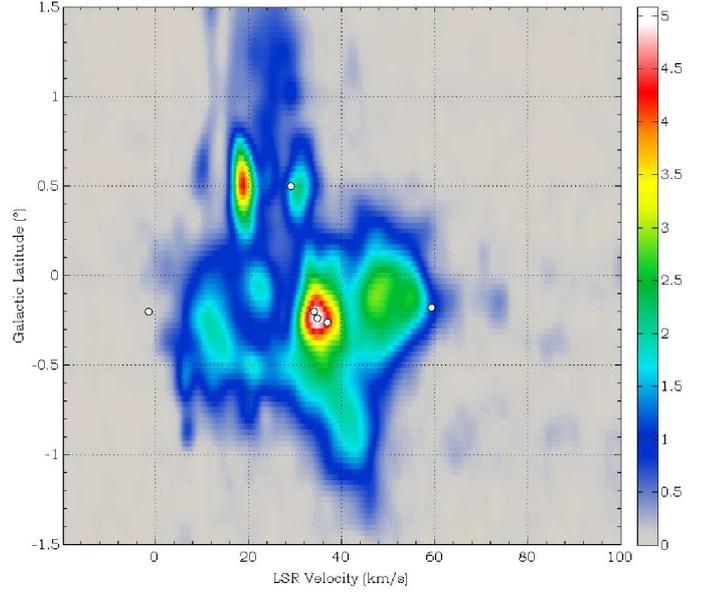} 
       \caption{Velocities of the water maser groups in the CO latitude-velocity
plot (white dots). The CO emission was integrated over a longitude
range from 12.625$\degr$ to 13.0$\degr$. The units of the color bar
are degrees Kelvin. The water masers in W33 and in G012.88+0.48 are associated
with clouds at 34.6 km s$^{-1}$ and 30.9 km s$^{-1}$, respectively.
The water masers at $-$1 km s$^{-1}$ (W33\,Main) and 58 km s$^{-1}$
(W33\,B) seem to be outliers.}                                           
	\label{LatvsVel} 
\end{figure}

The absolute parallaxes of G012.81--0.19, G012.88+0.48, G012.90--0.24,
and G012.90--0.26 are consistent with a parallax of 0.416 mas within
2$\sigma$, i.e. the distances to these masers are in accordance with the distance
to G012.68--0.18 (2.40 kpc).                                              
 
Comparing the coordinates of the W33 complex with a CO latitude-velocity
map from the survey of \citet{Dame2001} (Fig. \ref{LatvsVel}), we find that the 
complex is associated with a bright cloud at the same latitude with
a mean velocity of 34.6 km s$^{-1}$, which falls along the longitude-velocity locus
of the Scutum spiral arm.                                        
 
Since all water masers in the W33 complex are at a similar distance,
we conclude that the molecular clouds W33\,A, W33\,Main, and W33\,B 
are connected and belong to the same star forming complex. Our
parallax distance to this star forming cluster is a factor of 0.65 (= 2.4 kpc/3.7 kpc) 
smaller than the kinematic distance. Thus, the luminosity and mass of this complex
have been overestimated in the past by a factor of more than two
and should be revised to L $\sim$ 8$\cdot$10$^5$ L$_{\sun}$ and
M $\sim$ (0.8--8)$\cdot$ 10$^5$ M$_{\sun}$.                              
 
The spectral types of the stars in the W33\,Main cluster were determined
from the estimated number of Lyman continuum photons needed to ionize the
material in the surrounding $\ion{H}{II}$ region. Since the number
of photons depends on distance squared, the values have been
overestimated by a factor of more than two, changing the spectral
types by 1.5 points to later types, i.e., a star previous estimated
to be spectral type O6 should have a spectral type O7.5.
Thus, the star cluster in W33\,Main has spectral types ranging from
O7.5 to B1.5.                                                           
 

Our observations show maser detections in the W33 complex in three 
different velocity ranges: 
\begin{itemize}
\item 57 -- 63 km s$^{-1}$ in G012.68--0.18,
\item 33 -- 38 km s$^{-1}$ in G012.81--0.19, G012.90--0.24, G012.90--0.26,
\item $-$5 -- 0 km s$^{-1}$ in G012.81--0.19.
\end{itemize}
 These findings are confirmed 
by water and hydroxyl maser observations by \citet{Genzel1977}, 
\citet{Lada1981}, and \citet{Argon2000}. Ammonia 
observations with the Effelsberg 100 meter telescope \citep{Wienen2012} and from 
the HOPS survey \citep{Purcell2012} give line-of-sight
velocities of $\sim$55, $\sim$36, $\sim$36, and $\sim$37 km s$^{-1}$ for G012.68--0.18, 
G012.81--0.19, G012.90--0.24, and G012.90--0.26, respectively, in accordance with our
maser observations. On the other hand, \citet{Urquhart2008} and 
\citet{Chen2010} detected CO 
emission peaks at 35, 60, and 52 km s$^{-1}$ in G012.68--0.18, G012.81--0.19, and
G012.90--0.26. These results show that both velocity components at 
$\sim$36 and $\sim$58 km s$^{-1}$ are spread over the entire complex but the
emission at these velocities peaks in different regions. 
The ammonia observations of \citet{Wienen2012} show 
the velocity component at $\sim$58 km s$^{-1}$ also
at positions in the vicinity of W33 (e.g. G11.94--0.26, G12.74--0.10, G12.90--0.03).
This indicates that this velocity component is not unique to the inner W33 complex 
and probably has a different origin than internal motions (e.g. outflows) in the clouds. 
The velocity component at $\sim -$1 km s$^{-1}$ in G012.81--0.19 
was not detected in previous thermal line observations. The arrangement 
of the water maser spots in this velocity group (see Fig. \ref{G12.81spotmap}) 
might suggest that this velocity component is associated with the shock front of an 
outflow in the active cloud W33\,Main (Immer et al., in prep.) .     

Although water masers are often found at the base of bipolar outflows, 
the small radial velocity difference between the masing and the thermal gas in 
G012.81--0.19, G012.90--0.24, and G012.90--0.26 (see Table \ref{ParallaxResults}), 
the small radial velocity range of the masing gas and the proximity of the masers to 
submillimeter dust peaks suggests that the 3D motions of the water masers are similar 
to the motions of the associated central objects and not drastically 
altered by strong outflows. 
The only case where a strong outflow could affect the proper motion of the water 
maser without changing its radial velocity is if the outflow was aligned with the plane of the sky. 
However, the probability of outflows being in the plane of the sky in all three sources is small.
Thus, we assume that the average proper motion of the water maser spots in each 
source is similar to the motion of its central object. Under this assumption, we can estimate 
the cloud motions internal to the W33 complex from the motions of the central objects. 
In G012.68--0.18, we estimated the motion of the central object by fitting a model of 
expanding outflows to the proper motions of the maser spots. 
Since the maser spots in the 
velocity group around $-$1 km s$^{-1}$ in G012.81--0.19 have a very different radial velocity 
than the thermal gas in this cloud, their proper motions probably
do not reflect the motion of the central object of the W33\,Main cloud. They were thus excluded 
from the following calculations.
We averaged the motions of W33\,A (G012.90--0.24, G012.90--0.26), 
W33\,B (G012.68--0.18), and W33\,Main (G012.81--0.19 -- 34 km s$^{-1}$), 
yielding an estimate of the average proper motion of the complex: 
$\mu_x = -0.44 \pm 0.25$ mas yr$^{-1}$ and 
$\mu_y = -2.14 \pm 0.41$ mas yr$^{-1}$, 
corresponding to $-$5 km s$^{-1}$ and $-$24 km s$^{-1}$
in the eastward and northward directions, respectively (at the measured
distance of 2.40 kpc). To determine the cloud motions of W33\,A and 
W33\,B relative to the W33\,Main cloud, 
the motion of W33\,Main was subtracted from the motion of each cloud. 
Figure \ref{W33_ATLASGAL} shows the cloud motions of W33\,A and W33\,B 
in the reference frame of W33\,Main on the 870 $\mu$m dust emission 
map of the W33 complex.     

That W33\,A and W33\,Main have similar radial velocities suggests that 
most of the motion of W33\,A in the reference frame of W33\,Main is 
in the plane of the sky, tangentially to W33\,Main (total speed $\approx$ 17 
km s$^{-1}$). 
 
The difference in radial velocity between W33\,Main and W33\,B is $\sim$
22 km s$^{-1}$.
This large velocity component along the line of
sight coincidentially is equal to the velocity component in the plane of the
sky of 22 km s$^{-1}$ (relative to W33\,Main), yielding
a total speed of 31 km s$^{-1}$ for W33\,B relative to W33\,Main. 
In a forthcoming paper, we will compare the gravitational and kinetic energies 
of the clouds to determine if W33\,A and W33\,B are gravitationally bound to 
W33\,Main.

The massive star forming region G012.88+0.48 appears associated with a 
molecular cloud at a velocity of 30.9 km s$^{-1}$ (Fig. \ref{LatvsVel}). 
Thus, both cloud velocity and parallax distance locate this
source in the Scutum spiral arm.                                        
Since previous studies of G012.88+0.48 assumed a distance of 3.6 kpc,
its luminosity and mass have been overestimated 
by a factor of two and should be revised to L$\sim$1.3$\cdot$10$^5$
L$_{\sun}$ and M$\sim$1.3$\cdot$10$^3$ M$_{\sun}$.

\section{Conclusion}

Trigonometric parallax observations of water masers in the
G012.88+0.48 region and in the massive star forming complex W33 
(containing G012.68--0.18, G012.81--0.19, G012.90--0.24,
G012.90--0.26) yield distances which are all consistent with
$2.40^{+0.17}_{-0.15}$ kpc. A combination of our distance 
with CO observations of the Galactic plane by \citet{Dame2001} 
allows us to locate the W33 complex and G012.88+0.48 in the Scutum spiral arm. 
We show that W33 is a single star forming complex at about two-thirds the kinematic
distance of 3.7 kpc. Thus, the luminosity and mass estimates of this region should 
be lowered to L $\sim$ 8$\cdot$10$^5$ L$_{\sun}$ and 
M $\sim$ (0.8--8)$\cdot$ 10$^5$ M$_{\sun}$. 
The spectral types in the star cluster in W33\,Main should be changed by 1.5 points 
to later types, yielding spectral types ranging from O7.5 to B1.5.

The luminosity and mass values of G012.88+0.48 have also been overestimated in the past 
and revised values are 
L$\sim$1.3$\cdot$10$^5$ L$_{\sun}$ and M$\sim$1.3$\cdot$10$^3$ M$_{\sun}$.

\begin{acknowledgements} 
The authors would like to thank Koichiro Sugiyama for providing the 6.7 GHz 
methanol maser results of G012.88+0.48 from the Japanese VLBI Network survey. 
This work was partially funded by the ERC Advanced Investigator Grant GLOSTAR 
(247078). This work made use of the Swinburne University of Technology 
software correlator, developed as part of the Australian Major National Research 
Facilities Programme and operated under licence \citep{Deller2007}.
\end{acknowledgements}

\end{document}